\documentclass{aastex}
\usepackage{emulateapj5}

\shorttitle{$i'$-Dropout Overdensity in SDF}
\shortauthors{Ota et al.}

\begin{document}


\title{Overdensity of $i'$-Dropout Galaxies in the Subaru Deep Field:\\A Candidate Protocluster at $z\simeq 6$\altaffilmark{1,2}}

\author{Kazuaki Ota\altaffilmark{3}, Nobunari Kashikawa\altaffilmark{4,5}, Matthew A. Malkan\altaffilmark{6}, Masanori Iye\altaffilmark{4,5,7}, Tadashi Nakajima\altaffilmark{4}, Tohru Nagao\altaffilmark{8,9}, Kazuhiro Shimasaku\altaffilmark{7}, Poshak Gandhi\altaffilmark{3}}
\email{kz\_ota@crab.riken.jp}


\altaffiltext{1}{Based on data collected at Subaru Telescope, which is operated by National Astronomical Observatory of Japan.} 

\altaffiltext{2}{Some of the data presented herein were obtained at the W.M. Keck Observatory, which is operated as a scientific partnership among the California Institute of Technology, the University of California and the National Aeronautics and Space Administration. The Observatory was made possible by the generous financial support of the W.M. Keck Foundation.}

\altaffiltext{3}{Cosmic Radiation Laboratory, RIKEN, 2-1 Hirosawa, Wako-shi, Saitama 351-0198, Japan}

\altaffiltext{4}{National Astronomical Observatory of Japan, 2-21-1 Osawa, Mitaka, Tokyo, 181-8588, Japan}

\altaffiltext{5}{The Graduate University for Advanced Studies, 2-21-1 Osawa, Mitaka, Tokyo, 181-8588, Japan}

\altaffiltext{6}{Department of Physics and Astronomy, Box 951547, UCLA, Los Angeles, CA 90095, USA}

\altaffiltext{7}{Department of Astronomy, Graduate School of Science, University of Tokyo, 7-3-1 Hongo, Bunkyo-ku, Tokyo 113-0033, Japan}

\altaffiltext{8}{Research Center for Space and Cosmic Evolution, Ehime University, 2-5 Bunkyo-cho, Matsuyama 790-8577, Japan}

\altaffiltext{9}{Department of Physics, Graduate School of Science and Engineering, Ehime University, 2-5 Bunkyo-cho, Matsuyama 790-8577, Japan}



\begin{abstract}
We investigate the sky distribution of $z\sim6$ Lyman break galaxies selected as $i'$-dropouts having $i'-z'>1.45$  down to $z' \leq 26.5$ in the Subaru Deep Field (SDF). We discover 37 $i'$-dropouts clustered in a projected comoving $21.6 \times 21.6$ Mpc$^2$ region at $z=6$, showing a local density excess.  Carrying out follow-up spectroscopy, we identify four of them as Ly$\alpha$ emitters at $z=5.92$, 6.01, 6.03 and 6.03 (spread over a distance of 46.6 Mpc).  The number density of the cluster itself in SDF is $\sim 2.2\times 10^{-7}$ Mpc$^{-3}$, smaller than those of protoclusters (i.e., forming galaxy clusters) at $z\sim2$--5.7.  Also, the structure shows $\sim4$--21 times larger galaxy number density than those of $z\sim6$ galaxies in a general field.  It has a mass of $M\sim1.5^{+1.8}_{-0.5} \times 10^{15}M_{\odot}$, comparable to those of $z\sim0$--5 protoclusters.  Since the contamination of our sample by interlopers is estimated to be quite low, 5.9\%, most of the other unconfirmed $i'$-dropouts in the overdense region can be also $z\simeq 6$ galaxies.  Hence, it could be a candidate forming cluster at $z\simeq6$, representing a progenitor of galaxy clusters seen in the recent-day Universe.    
\end{abstract}




\section{Introduction}
Observations of high-$z$ star-forming galaxies such as Ly$\alpha$ emitters (LAEs) and Lyman break galaxies (LBGs) have been used to search for large-scale structures and primeval forming galaxy clusters---protoclusters---at high redshifts.  Starting from the discovery of protoclusters made of these galaxies at $z\sim2$--3 \citep{Malkan96,Pascarelle96,Giavalisco94,LeFevre96,Steidel98}, subsequent searches at higher redshifts have achieved remarkable success finding large-scale structures up to $z\sim4$--5.7 \citep{Overzier06,Venemans07,Shimasaku03,Kashikawa07,Malhotra05,Ouchi05}.  
Several attempts have been made to search for even higher-$z$ galaxy clustering targeting fields around $z\sim6$ quasars, but only two possible such system has been found \citep{Stiavelli05,Willott05,Zheng06,Ajiki06}.  

To  study galaxy clustering at even larger lookback time (i.e., $z\geq6$), 
we have selected $z\sim6$ galaxy candidates by the $i'$-band dropout technique, 
in the Subaru Deep Field \citep[SDF; $13^{\rm h}24^{\rm m}38.^{\rm s}9$, $+27\degr 29'25.''9$; an area of $\sim876$ arcmin$^2$]{Kashikawa04}, taking the full advantage of its extremely wide and deep optical images.  The present paper reports our finding of $i'$-dropout overdensity that could be a protocluster at $z\simeq 6$, and our follow-up spectroscopy of some of the $i'$-dropouts forming this overdense structure.  In \textsection 2, we describe the data used for our photometry as well as our $i'$-dropout sample selection.  The sky distribution of $i'$-dropouts in SDF is analyzed statistically in \textsection 3.  \textsection 4 presents the result of the follow-up spectroscopy.  Finally, discussion and conclusion are given in \textsection 5 and 6.  Throughout we adopt an $(\Omega_m, \Omega_{\Lambda}, h)=(0.3,0.7,0.7)$ cosmology and AB magnitude with $2''$ aperture unless otherwise specified.
  
\section{$i'$-Dropout Sample}
\subsection{Sample Selection}
We used official catalogs\footnote[8]{Available from http://step.mtk.nao.ac.jp/sdf/data/} of the objects detected in SDF, for which each of the deep $BVRi'z'$ images was taken by one pointing of the Subaru Prime Focus Camera \citep[Suprime-Cam]{Miyazaki02} on the 8.2m Subaru telescope.  All the images were convolved to have common seeing size of $0.''98$.  Source detection and photometry in each waveband were performed by using SExtractor version 2.1.6 \citep{BA96} and the limiting magnitudes in our $2''$ aperture at $3\sigma$ are $(B, V, R, i', z')=(28.45, 27.74, 27.80, 27.43, 26.62)$. The $2''$ aperture and total magnitudes were measured with MAG\_APER and MAG\_AUTO parameters, respectively.  From the objects detected down to $z'\leq 26.5$, we selected 158 $i'$-dropouts by using the criteria: $B>28.45$, $V>27.74$, $i' - z' > 1.45$.  Null detections ($<3\sigma$) in $B$ and $V$ were required since the flux of a LBG blueward of Ly$\alpha$ should be absorbed by the intergalactic medium.  Our $i' - z' > 1.45$ criterion picks up galaxies having Lyman breaks at $z\simeq 5$--7.1, based on calculations using population synthesis models \citep{BC03,madau95}.  This model calculation also shows that there could be weak remainig fluxes not completely absorbed by the intergalactic medium in the $R$-band wavelength of $z\sim6$ galaxies' spectra.  Hence, we did not require a null detection in $R$.  All the $i'$ magnitudes fainter than 28.62 ($1\sigma$ limits) were replaced by this value in application of the selection criteria. 

\subsection{Contamination from Interlopers \label{Contami}}
\begin{figure}
\plotone{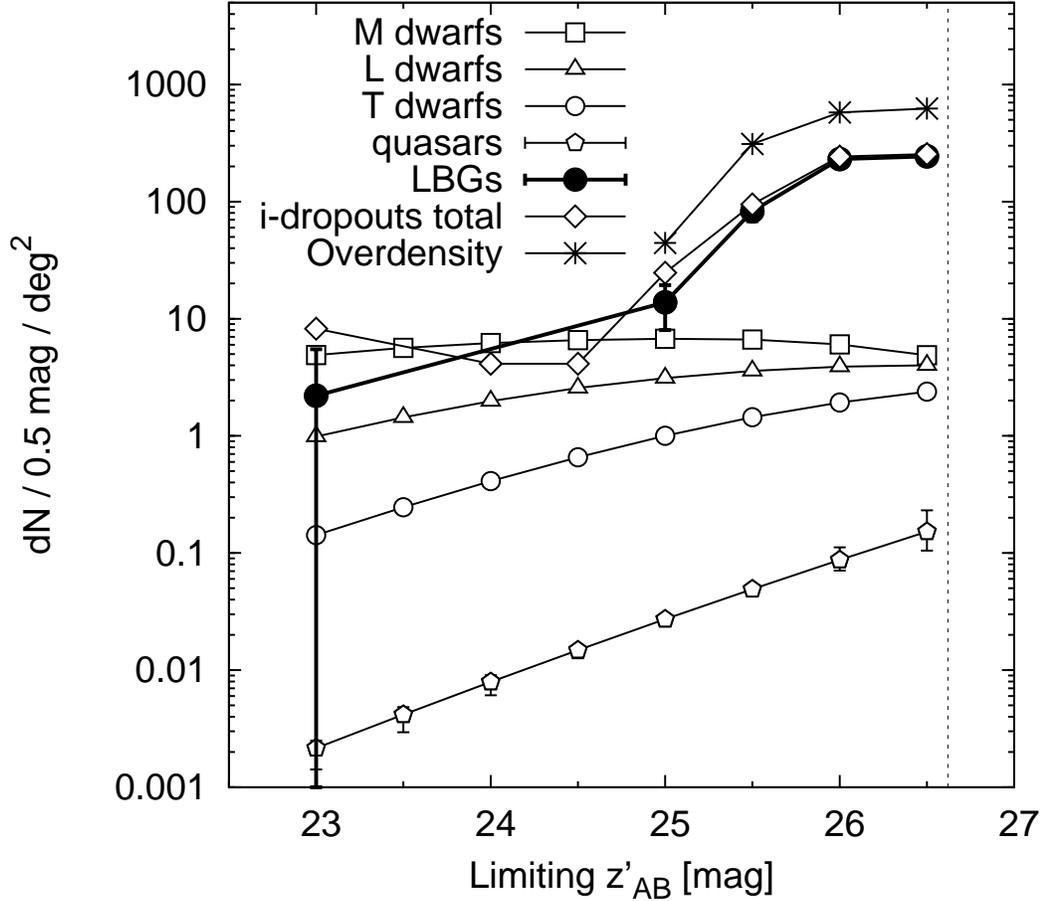}
\caption{Differential number counts of all the possible $i'$-dropout objects with $z'>23$ (total magnitude) in SDF.  They are shown in several different symbols and lines as labeled in the figure.  The vertical dotted line is the $3\sigma$ limiting magnitude of $z'=26
.62$.  All the vertical errors are Poissonian errors.  Poissonian errors for dwarfs and total $i'$-dropout counts are omitted for clarity.  The count of $z\sim6$ LBGs is the difference between $i'$-dropouts total and the sum of M/L/T dwarf star and $z\sim6$ quasar.  $\sim91$\% of the $i'$-dropouts have $z'> 25$ where the contamination rates of dwarfs, EROs and quasars are only 5.9\%, 0\% and 0.12\%, respectively.  The surface density of $i'$-dropouts in the $9' \times 9'$ overdense region (See \textsection \ref{ClusterDensity} for the size definition) is also shown by asterisks.  It is a factor of 2--3 times larger than the surface density of $i'$-dropouts in the entire SDF field.\label{NumbCount}}
\end{figure}

The problem in selecting $z\sim6$ galaxies without near-infrared images is contamination.  Three different types of interlopers could satisfy $i'-z'>1.45$; dwarf stars with spectral type later than M7 \citep{Hawley02}, $z\sim 1$--2 old elliptical galaxies called Extremely Red Objects (EROs) with 4000\AA~Balmer break in $z'$-band and $z\sim6$ quasars with Ly$\alpha$ absorption.  We evaluated fractions of these interlopers in our sample using the same method as \citet{Ota05}: using a star count model \citep{Nakajima00}, old elliptical ERO samples actually observed in Subaru/{\it XMM-Newton} Deep Survey \citep{Miyazaki03} and Hubble Ultra Deep Field (UDF) \citep{Yan04}, and a quasar luminosity function \citep{YanWin04}.  The calculated number counts of each interloper as well as $z\sim6$ LBG are shown in Figure \ref{NumbCount}.  Most ($\sim 91$\%) of the $i'$-dropouts in SDF are fainter than $z'=25$.  The possible fractions of dwarfs, EROs and quasars in $z'>25$ are only 5.9\%, 0\% and 0.12\%, respectively.  If we consider cosmic variance of ERO number density \citep{Some04} and Poisson error for 0 \citep{Geh86}, the upper limit of ERO fraction could be $\lesssim 16$\% in SDF survey area if these EROs satisfy $i'-z'>1.45$.  
However, \citet{Malhotra05} also confirmed that no ERO shows $i'-z'>1.3$ color (See their \textsection 3 and Fig.1) based on their spectroscopy of $i$-dropouts in UDF.  Hence, we ignore the ERO contamination in the subsequent analyses and discussions.  The low contamination rate estimated here is consistent with the fact that our follow-up spectroscopy did not reveal any candidates to be interlopers (See \textsection \ref{spec}).        

\section{Sky Distribution}
\subsection{Discovery of A Local Surface Overdensity}
\begin{figure}
\plotone{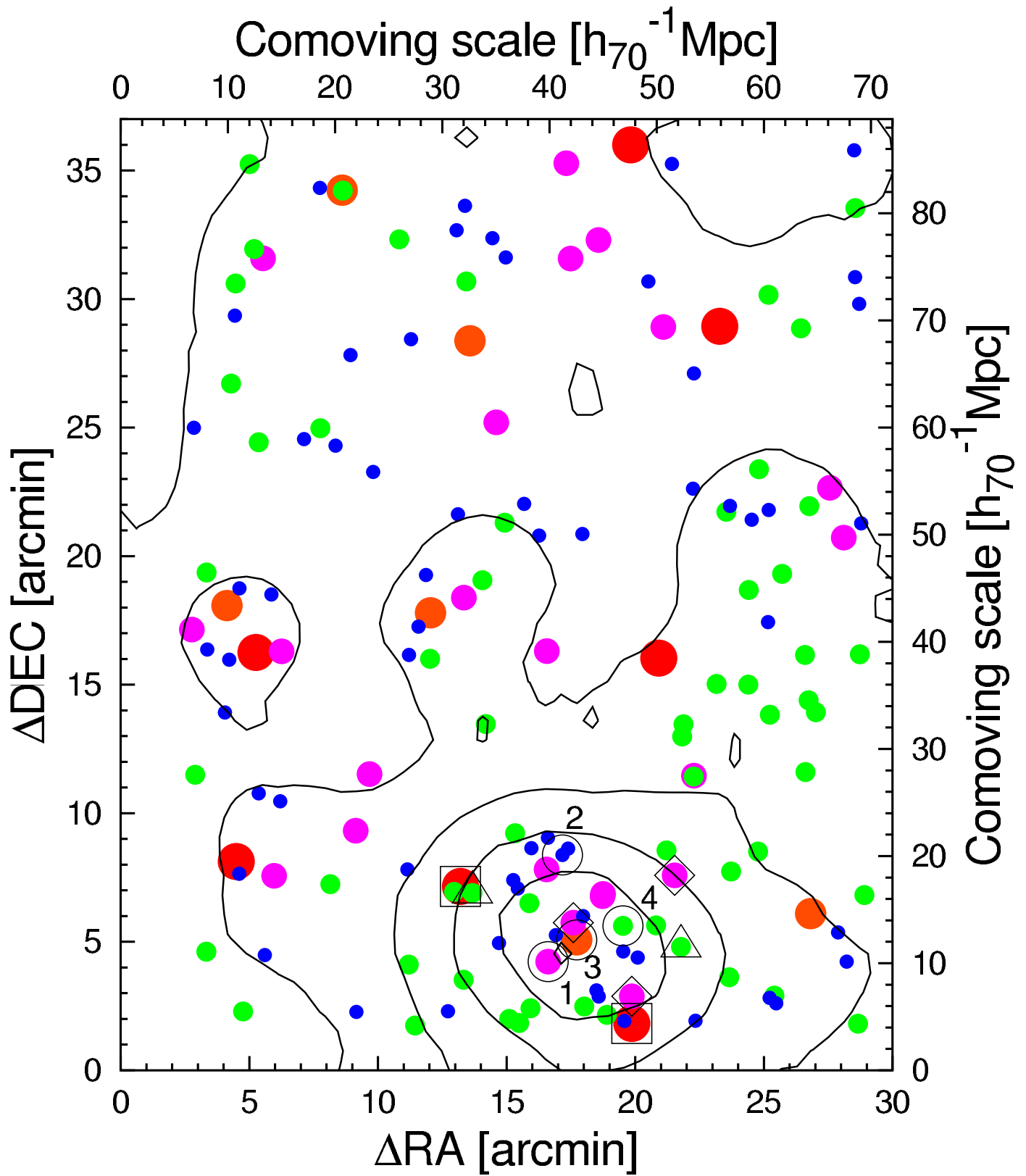}
\caption{Sky distribution of $i'$-dropouts in the Subaru Deep Field.  They are filled circles with sizes proportional to their $z'$-band total magnitudes (huge red: $z'\leq24.5$, very large orange: $24.5<z'\leq25$, large magenta: $25<z'\leq25.5$, medium green: $25.5<z'\leq26$, small blue: $26<z'\leq26.5$).  North is up, and east is to the left.  The comoving scale projected to $z=6$ is also shown along the axes.  Solid lines are density contours of $\delta _{\Sigma}=0$, 0.2, 0.4, 0.6, 0.8 and 1.0 (See \textsection 3.1 for the definition of $\delta _{\Sigma}$). Four spectroscopically confirmed Ly$\alpha$ emitters at $z=5.92$, 6.01, 6.03 and 6.03 are encircled and labeled with numbers 1--4 as in Table \ref{LAEProperty} and Figure \ref{Spectra}.  They are located in the overdense region to the south.  Two $i'$-dropouts enclosed by squares are extremely bright ($z'=21.88$ and 22.16) and might be $z\simeq6$ quasars physically associated with $z\simeq6$ galaxies or possibly dwarf stars.  Three and two $i'$-dropouts enclosed by diamonds and triangles in the overdense region are more plausible candidates for $z\sim6$ LBGs selected by using $z_R$ filter by \citet{Shimasaku05} and by the narrowband NB816-dropout technique by \citet{Shioya05}.\label{SkyDistribution}}
\end{figure}

The sky distribution of 158 $i'$-dropouts in SDF is shown in Figure \ref{SkyDistribution}. 
A very interesting region in the south shows a significant overdensity of these $i'$-dropouts which may be a protocluster at $z \sim 6$.  
To verify the significance of this compared to other regions quantitatively, we measured the degree of density excess, $\sigma_{\Sigma20}$, the overdensity of $i'$-dropouts in a circle of 20 Mpc comoving radius at $z=6$ defined as $\sigma_{\Sigma20}^2 \equiv [\langle (\Sigma - \bar{\Sigma})^2 \rangle - \bar{\Sigma}]/\bar{\Sigma}^2$ \citep{Peebles80}.  Here, $\Sigma$ and $\bar{\Sigma}$ are the number of objects in the circle and the mean of $\Sigma$ values measured in circles at 2000 randomly chosen positions in the $z'$ image of SDF, respectively.  We obtained $\sigma_{\Sigma20} = 0.63$ with $3\sigma$ significance for the region of our interest.  This value is much higher than those of any other places in SDF, and comparable to that of $z \sim 5.7$ protoclusters found by \citet{Ouchi05}.  This area also shows the highest density contrast: $\sigma_{\Sigma} \equiv (\Sigma - \bar{\Sigma})/\bar{\Sigma} = 3.6$ with 5$\sigma$ excess measured in a comoving 8 Mpc circle.  This value is comparable to those for a $z \sim 4.8$ structure found by \citet{Shimasaku03,Shimasaku04} and $z \sim 5.7$ protoclusters by \citet{Ouchi05}.  We found 56 and 22 $i'$-dropouts in 20 Mpc and 8 Mpc circles, respectively.  


The overdense region contains two extremely bright ($z'=21.88$ and 22.16) $i'$-dropouts that might be $z \sim 6$ quasars or dwarf stars as seen in Figure \ref{SkyDistribution}.  They could be part of a system consisting of LAEs physically associated with one or two quasars at $z\simeq 6$, although these bright $i'$-dropouts are more plausibly dwarf stars.  

In the SDF, $z\sim6$ galaxy candidates had been also selected by two different techniques.  \citet{Shimasaku05} made $z_B$ and $z_R$ filters splitting the $z'$ waveband at 9500\AA~and selected out $z=5.9 \pm 0.3$ LBG candidates in the $i'-z_R$ vs. $z_B -z_R$ color plane down to $z_R=25.4$.  Meanwhile, \citet{Shioya05} detected $z=5.8$--6.5 LBG candidates (NB816-dropouts) using the $i'-z'$ vs. NB816 $-z'$ color diagram down to $z'=26.07$, where NB816 is a narrowband filter centered at 8150\AA~with FWHM of 120\AA.  Our overdensity region includes three such $z_R$-detected objects out of 14 $i'$-dropouts down to $z'=26.22$ (total mag) assuming it is comparable to $z_R=25.4$ ($z' \sim z_R+0.82$ on average; see Table 2 in \citet{Shimasaku05}) and two NB816-dropouts out of 14 $i'$-dropouts down to $z'=26.07$.  They are regarded as more plausible galaxy candidates at $z\sim6$.  Note that since their candidate selections adopt null detection in $R$-band ($R<2$--$3\sigma$), we also limited our sample to $R<2$--$3\sigma$ in the comparison.



\section{Follow-up Spectroscopy\label{spec}}
To obtain information about the physical clustering and to check the validity of our sample selection, we obtained spectra of 20 $i'$-dropouts in the overdense region with Faint Object Camera and Spectrograph \citep[FOCAS]{Kashikawa02} on the Subaru Telescope and the DEIMOS spectrograph \citep{Faber03} on the Keck II Telescope.  Spectroscopy of these $i'$-dropouts is summarized in Table \ref{SpecStatus}, and in
\citet{Kashikawa06}, \citet{Iye06} and \citet{Ota08}.

\begin{figure}
\plotone{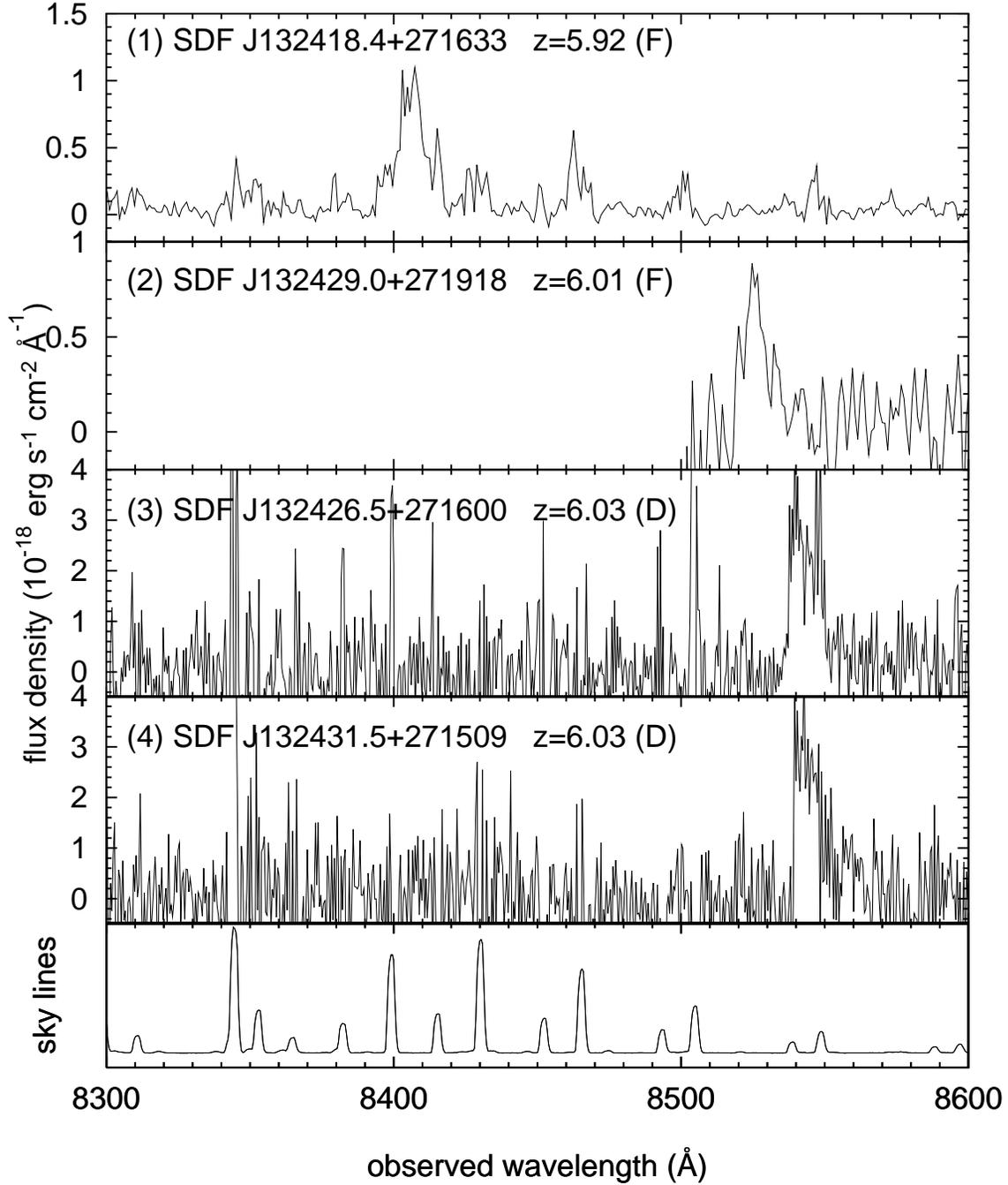}
\caption{Spectra of four confirmed $z\simeq6$ Ly$\alpha$ emitters in the overdense region as well as OH skylines taken with Subaru/FOCAS in the same wavelength region.  Spectra obtained using Subaru/FOCAS and Keck/DEIMOS are labeled (F) and (D), respectively.  The labels of (1)--(4) correspond to those in Figure \ref{SkyDistribution} and Table \ref{LAEProperty}.\label{Spectra}}
\end{figure}

We identified four $i'$-dropouts as LAEs at similar redshifts $z=5.92$, 6.01, 6.03 and 6.03.  The emission lines found in their spectra are asymmetric with their blue side absorbed and have weighted skewness of $S_W=3.43$--17.9\AA, comparable to the values of $z=5.7$ and 6.6 LAEs, $S_w^{z=5.7}=3.33$--20.66\AA~and $S_w^{z=6.6}=3.12$--16.34\AA~(See Table \ref{LAEProperty} and \citet{Shimasaku06} and \citet{Kashikawa06}).
Since each of the four spectra shows only one emission line, the line is not H$\beta$, [OIII]4959\AA, [OIII]5007\AA, H$\alpha$, [SII]6717\AA~or [SII]6731\AA.  Also, the emission lines we found are not the $z=1.6$ [OII] doublet with 7\AA~separation (3726\AA~and 3727\AA~at rest frame) because we actually resolved even closer (5--6\AA) skylines in the same wavelength range.  In addition, the objects are not $z\simeq 6$ active galactic nuclei, since we did not find N$_{\rm V}$ 1240\AA~lines at 8680\AA.  Also, FWHMs of the 4 confirmed emssion lines are 367--464 km s$^{-1}$ and much smaller than the average FWHM of $z\sim6$ quasars' Ly$\alpha$ lines, $\sim6,000$ km s$^{-1}$ \citep{Fan04}.  Hence, we consider all the emission lines we detected to be Ly$\alpha$ of LAEs.  The photometric and spectroscopic properties of four confirmed LAEs as well as their spectra, are presented in Table \ref{LAEProperty} and Figure \ref{Spectra}.  Their sky distribution is shown in Figure \ref{SkyDistribution}. 

Spectra of 15 out of 16 unconfirmed $i'$-dropouts were obtained during the 2005 run whose main targets were other  brighter objects, and thus integration times were rather short (only 1.25--3.75 hours) for our faint ($z'=25.5$--26.5) $i'$-dropouts.  Hence, if the Ly$\alpha$ lines are absent or overlapped with OH-airglow lines, identification is much harder.  Thus, we cannot completely deny the possibility of these unconfirmed $i'$-dropouts being galaxies at $z\simeq6$.  None of the 20 spectroscopic targets turned out to be low-$z$ emission-line galaxies.  Hence, the possible contamination of low-$z$ galaxies in our photometric $i'$-dropout sample seems very rare.        

\section{Discussion}
Analyses of the overdense region implies that it might be a $z\simeq6$ protocluster.  In the following discussion, we estimate the basic properties of the overdense region and compare them with those of $z<6$ protoclusters and field galaxies to see if it can be a $z\simeq6$ protocluster that is a progenitor of galaxy clusters seen in the present-day universe.

\subsection{Number Density of the Cluster Itself \label{ClusterDensity}}
Since we have probed the comoving depth of 887 Mpc (corresponding to $z\simeq 5$--7.1 of $i'$-dropout selection function) and area of SDF (876 arcmin$^2$), our survey volume is approximately $4.5\times10^6$ Mpc$^3$.  Since we found only one overdensity in this volume, its number density is estimated to be $2.2\times 10^{-7}$ Mpc$^{-3}$.  This is smaller than those of protoclusters at lower-$z$; e.g., $2.2\times 10^{-6}$ Mpc$^{-3}$ at $z=5.7$ \citep{Ouchi05}, $7.1\times 10^{-6}$ Mpc$^{-3}$ at $z=4.86$ \citep{Shimasaku03}, $\sim6.0\times 10^{-6}$ Mpc$^{-3}$ at $z=2$--5 \citep{Venemans07}, $\sim3.0\times 10^{-6}$ Mpc$^{-3}$ at $z\sim3$ \citep[converted to our cosmology]{Steidel98}.  The smaller number density might be because forming clusters are not virialized yet and difficult to identify at higher redshifts as implied by hierarchical clustering scenario.   

\subsection{Galaxy Number Density \label{NumDen}}
If we define the border of the overdense region as the place where the density contrast drops to the half of unity $\delta_{\Sigma}\sim0.5$, its size is  $\sim 9' \times 9'$ ($21.6 \times 21.6$ comoving Mpc$^2$ at $z=6$), approximately within $\rm \Delta RA=13'$--$22'$ and $\rm \Delta DEC=1'$--$10'$ in Figure \ref{SkyDistribution}.  There are a total of 37 $i'$-dropouts in it.  Four of them are identified as LAEs, of which one has $z'<25$.  Also, two bright objects might be dwarf stars or quasars.  Since the contamination rate of dwarfs in $z'>25$ is expected to be 5.9\% (See \textsection \ref{Contami}), 32 out of the remaining 34 $i'$-dropouts in $z'>25$ can be $z\sim6$ galaxies.  Hence, we assume that the overdensity consists of 33 galaxies at $z\simeq6$.  If we assume the line-of-sight scale of the overdensity corresponds to $z=5.92$--6.03 of confirmed LAEs (46.6 comoving Mpc), its volume would be $2.2\times 10^4$ Mpc$^3$.  This gives the galaxy number density of $n_{\rm cluster} \sim 1.5^{+1.4}_{-1.0} \times 10^{-3}$ Mpc$^{-3}$.  The uncertainty is large since it includes Poisson errors as well as large cosmic variance due to the small volume.  Cosmic variance is estimated in the same way as in \citet{Ota08}.  For comparison, we calculated the number density of $z\sim6$ galaxies in the general field and obtained $n_{\rm field} \sim 4.0^{+1.6}_{-1.4} \times 10^{-4}$ Mpc$^{-3}$ (same uncertainty as above) by integrating the best-fit Schechter luminosity function (LF) \citep{Schechter76} of $i$-dropouts derived by \citet{Bouwens06} down to $M_{\rm 1500,AB}=-20.2$ (corresponding to our detection limit $z'=26.5$).  Thus, the overdensity region has about four times ($n_{\rm cluster}/n_{\rm field} \sim 3.8^{+7.3}_{-2.6}$) higher surface density than galaxies in a field.  Note that this factor and $n_{\rm cluster}$ are lower limits because the incompleteness regarding $i'$-dropout selection is not taken into account.  Hence, those values could be larger if the incompleteness were corrected.         

\subsection{Star Formation Rate}
We estimate the average UV star formation rate (SFR) of the overdense region, excluding two bright $i'$-dropouts that might be dwarfs or quasars and assuming that all the rest of 33 $i'$-dropouts (corrected for contamination) are galaxies at $z=6$.  Their rest frame UV continuum luminosities, $L_{\nu}$, are calculated from $z'$-band total magnitudes and converted to SFR at 1500\AA, using the formalism of \citet{Madau98}: $L_{\nu}=8.0 \times 10^{27} {\rm SFR}$ erg s$^{-1}$ Hz$^{-1}$.  This yields the average SFR of $11.5 M_{\odot}$yr$^{-1}$.  Meanwhile, the average SFR of 116 $i'$-dropouts with the same magnitude cuttoff of $z'>24.5$ (total mag) outside the overdense region in SDF is $12.0 M_{\odot}$yr$^{-1}$.  This implies that star formation activity in the the overdense region does not differ much from that in a general field.    



 

\subsection{Mass Estimate \label{mass}}
Now that the galaxy overdensity $\delta_{\rm gal} \equiv n_{\rm cluster}/n_{\rm field} -1\sim 2.8^{+7.3}_{-2.6}$ is known, we can estimate the mass of the overdensity region with $M=\overline{\rho}V(1+\delta_m)$ \citep{Steidel98,Venemans05}.  Here, $\overline{\rho}=\Omega_m(3H_0/8 \pi G)^2=4.1 \times 10^{10}M_{\odot}$~Mpc$^{-3}$ is the mean matter density of the Universe with gravitational constant $G$.  $V=2.2\times 10^4$ Mpc$^3$ is the volume of the candidate protocluster.  $\delta_m$ is the mass overdensity related to the bias parameter $b$ with $1+b\delta_m=C(1+\delta_{\rm gal})$.  $C=[1+f-f(1+\delta_m^{1/3})]$, where $f\simeq1$ in our cosmology, corrects the volume for redshift space distortion due to peculiar motions of protocluster galaxies \citep{Steidel98,Venemans05}.  Using $\delta_{\rm gal}\sim 2.8^{+7.3}_{-2.6}$ and $b=3.4 \pm 1.8$ suggested by \citet{Ouchi05}, we estimate that $\delta_m \sim 0.62^{+2.0}_{-0.6}$ and thus $M\sim1.5^{+1.8}_{-0.5} \times 10^{15}M_{\odot}$.  Our estimated mass is comparable to the mass of the $z=4.1$ protocluster, (1--$2)\times 10^{15}M_{\odot}$, estimated by \citet{Venemans02}, that of the $z\sim3$ LBG concentration, (0.98--$1.4)\times 10^{15}M_{\odot}$ by \citet{Steidel98}, those of $z=2$--5 protoclusters, (2--$9)\times 10^{14}M_{\odot}$, by \citet{Venemans07}, and that of a $z=4.86$ protocluster, $3\times 10^{14}M_{\odot}$, by \citet{Shimasaku03}.  According to the large sample statistics (152--170 clusters) of nearby ($z\leq0.15$) galaxy clusters studied by \citet{Girardi98a,Girardi98b}, our overdensity mass is also comparable to their masses.  Moreover, the estimated mass and number density of the overdensity measured in \textsection \ref{ClusterDensity} are consistent with mass functions derived from large samples of $\sim72$--300 $z\sim0.1$--0.2 nearby galaxy clusters \citep{Girardi98b,Bahcall03,Rines07}.  Hence, like other lower-$z$ protoclusters, it might be an ancestor of a cluster of galaxies seen in the present-day Universe.

\section{Conclusion}
The number density, number density of member galaxies and mass of the $i'$-dropout overdensity in SDF imply that there exist forming clusters---protoclusters---at a redshift as high as $z\simeq 6$.  However, the result presented here is based on photometry and a few spectroscopically identified LAEs at $z=5.92$--6.03.  Hence, the overdensity is still a "candidate" protocluster.  Further follow-up deep spectroscopy of the unconfirmed $i'$-dropouts will reveal if it is a real $z\simeq 6$ protocluster and constrain its physical properties.

\acknowledgments
We thank staff at Subaru and Keck Telescope for their expert support.  We express our gratitude to the SDF team for making such invaluable imaging data public.  This search was supported by the Japan Society for the Promotion of Science through Grant-in-Aid for Scientific Research 19540246.  K.O. acknowledges fellowship support from the Special Postdoctoral Researchers Program at RIKEN.  




\clearpage





\clearpage

\begin{table}
\begin{center}
\footnotesize
\caption{Status of the follow-up spectroscopy of $i'$-dropouts in the overdense region\label{SpecStatus}}
\begin{tabular}{llrccccccccc}\tableline\tableline
Ob     & ID    & date                       & seeing   & exposure & I & grating     & $R$  & wavelength & filter & slit & standard stars$^e$\\
       &       & (HST)                      & ($''$)   & (seconds) &   & (mm$^{-1}$) &      & (nm)      &           &($''$)&         \\\tableline
 2$^a$ & 1     & 24--27 Apr 2004$^c$        & 0.4--0.8 & 16200    & F & 300         & 1300 & 540--1000 & $O58$     & 0.6  & Hz 44, Feige 34\\
 8$^a$ & 0     & 14--15 May, 1 Jun 2005$^d$ & 0.5--1.0 & 13500    & F & 175         & 1600 & 830--1000 & SDSS $z'$ & 0.8  & Hz 44, Fiege 34, Feige 110\\ 
 4     & 1     & 14--15 May, 1 Jun 2005$^d$ & 0.5--1.0 & 9000     & F & 175         & 1600 & 830--1000 & SDSS $z'$ & 0.8  & Hz 44, Fiege 34, Feige 110\\ 
 5     & 0     & 14--15 May 2005$^d$        & 0.5--1.0 & 4500     & F & 175         & 1600 & 830--1000 & SDSS $z'$ & 0.8  & Fiege 34, Feige 110\\ 
 2     & 2$^b$ & 23--24 Apr 2004$^c$        & 0.6--1.0 & 9000     & D & 830         & 3600 & 500--1000 & GG495     & 1.0  & Feige 110, BD+28$^{\circ}$4211\\\tableline
\end{tabular}
\tablenotetext{}{NOTE: Columns in the table show the number of $i'$-dropouts spectroscopically observed (Ob), the number of $i'$-dropouts identified (ID), the dates of observation, typical seeing size during the observation, exposure time, the instrument (I) used (F for Subaru/FOCAS and D for Keck/DEIMOS, respectively.} 
\tablenotetext{a}{One of them was observed in both dates.}
\tablenotetext{b}{One of them was identified by \citet{Nagao05,Nagao07}.}
\tablenotetext{c}{Spectra were obtained in the observations by \citet{Kashikawa06}}
\tablenotetext{d}{Spectra were obtained in the observations by \citet{Iye06} and \citet{Ota08}.}
\tablenotetext{e}{See \citet{Oke90} and \citet{Hamuy94} for details.}
\end{center}
\end{table}


\clearpage
\begin{table}
\begin{center}
\scriptsize
\caption{Properties of four confirmed $z \simeq 6$ Ly$\alpha$ emitters in the overdense region\label{LAEProperty}}
\begin{tabular}{lccccccccc}\tableline\tableline
Object and Coordinate       & $z$ & $i'$ & $z'$ & $S_w$ & $F(\rm Ly\alpha)$ & $L(\rm Ly\alpha)$ & $SFR(\rm Ly\alpha)$ & $FWHM$ & $EW_0(\rm Ly\alpha)$\\
                            &      & (mag) & (mag) & (\AA) & (10$^{-17}$erg s$^{-1}$ cm$^{-2}$) & (10$^{43}$erg s$^{-1}$) & (M$_{\odot}$ yr$^{-1}$) & (\AA) & (\AA)\\\tableline
(1) SDF~J132418.4+271633$^{a}$   & 5.92 & 27.30    & 25.74 & $3.43\pm1.07$ & $1.3\pm0.2$  & $0.49\pm0.07$ & $4.43\pm0.6$ & 13.5 & 36 \\
(2) SDF~J132429.0+271918$^{a}$   & 6.01 & $>$28.62 & 26.29 & $8.52\pm1.39$ & $0.84\pm0.5$ & $0.34\pm0.19$ & $3.06\pm1.7$ & 11.2 & 30  \\ 
(3) SDF~J132426.5+271600$^{b,c}$ & 6.03 & 27.43    & 25.36 & $5.12\pm1.47$ & $3.6\pm0.3$  & $1.5\pm0.1$   & $13.7\pm0.9$ & 13.2 & 84\\
(4) SDF~J132431.5+271509$^{b}$   & 6.03 & 27.83    & 25.88 & $17.9\pm0.32$ & $4.3\pm0.3$  & $1.7\pm0.1$   & $15.6\pm0.9$ & 14.2 & 81 \\\tableline
\end{tabular}
\tablenotetext{}{NOTE: Units of coordinate are hours: minutes: seconds (right ascension) and degrees: arcminutes: arcseconds (declination) using J2000.0 equinox.  $i'$ and $z'$ are $2''$ aperture magnitudes.  Magnitude is replaced by its $1\sigma$ limit if it is fainter than the limit.  The indicator of a line profile asymmetry, weighted skewness $S_w$, is calculated using the definition  in \citet{Shimasaku06} and \citet{Kashikawa06}.  The Ly$\alpha$ line fluxes, $ F$(Ly$\alpha$) measured in the spectra are used to calculate the Ly$\alpha$ line luminosities, $L$(Ly$\alpha$).  Then, $L$(Ly$\alpha$) is converted into the corresponding star formation rates, $SFR$(Ly$\alpha$), using the relation, $SFR({\rm Ly}\alpha)=9.1\times10^{-43}L({\rm Ly}\alpha)M_{\odot}{\rm yr}^{-1}$, derived from Kennicutt's equation \citep{Kennicutt98} with the case B recombination theory. The $FWHM$ of each Ly$\alpha$ line is measured by fitting a Gaussian profile.  The rest frame equivalent widths of Ly$\alpha$ lines $EW_0(\rm Ly\alpha)$ are calculated in the same manner as in \citet{Nagao07}.} 
\tablenotetext{a}{Confirmed by Subaru/FOCAS.}
\tablenotetext{b}{Confirmed by Keck/DEIMOS.}
\tablenotetext{c}{Identified by \citet{Nagao05,Nagao07}.}
\end{center}
\end{table}

\clearpage



\begin{thebibliography}{ }
\bibitem[Ajiki et al.(2006)]{Ajiki06} Ajiki, M. et al.  2006, \pasj, 58, 499
\bibitem[Bahcall et al.(2003)]{Bahcall03} Bahcall, N. A.  et al. 2003, \apj, 585, 182
\bibitem[Bertin \& Arnouts(1996)]{BA96} Bertin, E., \&  Arnouts, S.  1996, \aap, 117, 393
\bibitem[Bouwens et al.(2006)]{Bouwens06} Bouwens, R.J., Illingworth, G.D., Blakeslee, J.P., \& Franx, M.  2006, \apj, 653, 53
\bibitem[Bruzual \& Charlot(2003)]{BC03} Bruzual, A.G., \& Charlot S.  2003, \mnras, 344, 1000  
\bibitem[Faber et al.(2003)]{Faber03} Faber, S.M. et al. 2003, Proc. SPIE, 4841, 1657
\bibitem[Fan et al.(2004)]{Fan04} Fan, X. et al.  2004, \aj, 128, 515
\bibitem[Gehrels~(1986)]{Geh86} Gehrels, N.,  1986, \apj, 303, 336
\bibitem[Giavalisco et al.(1994)]{Giavalisco94} Giavalisco, M., Steidel, C. C., \& Szalay, A. S. 1994, \apjl, 425, L5
\bibitem[Girardi et al.(1998a)]{Girardi98a} Girardi, M., Giuricin, G., Mardirossian, F., Mezzetti, M., \& Boschin, W.  1998, \apj, 505, 74 
\bibitem[Girardi et al.(1998b)]{Girardi98b} Girardi, M., Borgani, S., Giuricin, G., Mardirossian, F., \& Mezzetti, M.  1998, \apj, 506, 45
\bibitem[Hamuy et al.(1994)]{Hamuy94} Hamuy, M., Suntzeff, N.B., Heathcote, S.R., Walker, A.R., Gigoux, P., \& Phillips, M.M.  1994, \pasp, 106, 566
\bibitem[Hawley et al.(2002)]{Hawley02} Hawley, S.L. et al.  2002, \aj, 123, 3409
\bibitem[Iye et al.(2006)]{Iye06} Iye, M. et al.  2006, \nat, 443, 186
\bibitem[Kashikawa et al.(2002)]{Kashikawa02} Kashikawa, N. et al.  2002, \pasj, 54, 819
\bibitem[Kashikawa et al.(2004)]{Kashikawa04} Kashikawa, N. et al.  2004, \pasj, 56, 1011
\bibitem[Kashikawa et al.(2006)]{Kashikawa06} Kashikawa, N. et al.  2006, \apj, 648, 7
\bibitem[Kashikawa et al.(2007)]{Kashikawa07} Kashikawa, N., Kitayama, T., Doi, M., Misawa, T., Komiyama, Y., \& Ota, K. 2007, \apj, 663, 765
\bibitem[Kennicutt(1998)]{Kennicutt98} Kennicutt, R.C., Jr.,  1998, \araa, 36, 189
\bibitem[Le F\'{e}vre et al.(1996)]{LeFevre96} Le F\'{e}vre, O., Deltorn, J. M., Crampton, D., \& Dickinson, M.  1996, \apjl, 471, L11
\bibitem[Madau(1995)]{madau95} Madau, P.  1995, \apj, 441, 18
\bibitem[Madau et al.(1998)]{Madau98} Madau, P., Pozzetti, L., \& Dickinson, M.  1998, \apj, 498, 106
\bibitem[Malhotra et al.(2005)]{Malhotra05} Malhotra, S. et al. 2005, \apj, 626, 666
\bibitem[Malkan et al.(1996)]{Malkan96} Malkan, M. A., Teplitz, H., \& McLean, I. S.  1996, \apjl, 468, L9
\bibitem[Miyazaki et al.(2002)]{Miyazaki02} Miyazaki, S. et al.  2002, \pasj, 54, 833
\bibitem[Miyazaki et al.(2003)]{Miyazaki03} Miyazaki, M. et al.  2003, \pasj, 55, 1079
\bibitem[Nagao et al.(2005)]{Nagao05} Nagao, T. et al.  2005, \apj, 634, 142
\bibitem[Nagao et al.(2007)]{Nagao07} Nagao, T. et al.  2007, \aap, 468, 877
\bibitem[Nakajima et al.(2000)]{Nakajima00} Nakajima, T. et al.  2000, \aj, 120, 2488
\bibitem[Oke(1990)]{Oke90} Oke, J.B. 1990, \aj, 99, 1621
\bibitem[Ota et al.(2005)]{Ota05} Ota, K., Kashikawa, N., Nakajima, T., \& Iye, M.  2005, JKAS, 38, 179
\bibitem[Ota et al.(2008)]{Ota08} Ota, K. et al.  2008, \apj, 677, 12
\bibitem[Ouchi et al.(2005)]{Ouchi05} Ouchi, M. et al.  2005, \apjl, 620, L1
\bibitem[Overzier et al.(2006)]{Overzier06} Overzier, R.A. et al.  2006, \apj, 637, 58
\bibitem[Pascarelle et al.(1996)]{Pascarelle96} Pascarelle, S. M., Windhorst, R. A., Driver, S. P., Ostrander, E. J., \& Keel, W. C.  1996a, \apjl, 456, L21
\bibitem[Peebles~(1980)]{Peebles80} Peebles, P. J. E. 1980, The Large-Scale Structure of the Universe (Princeton: Princeton University Press)
\bibitem[Rines et al.(2007)]{Rines07} Rines, K., Diaferio, A., \& Natarajan, P.  2007, \apj, 657, 183
\bibitem[Reiprich \& B$\ddot{\rm o}$hringer(2002)]{Reiprich2002} Reiprich, T.H., \& B$\ddot{\rm o}$hringer, H.  2002, \apj, 567, 716 
\bibitem[Schechter (1976)]{Schechter76} Schechter, P. et al.  1976, \apj, 203, 297
\bibitem[Shimasaku et al.(2003)]{Shimasaku03} Shimasaku, K. et al.  2003, \apj, 586, L111
\bibitem[Shimasaku et al.(2004)]{Shimasaku04} Shimasaku, K. et al.  2004, \apj, 605, L93
\bibitem[Shimasaku et al.(2005)]{Shimasaku05} Shimasaku, K., Ouchi, M., Furusawa, H., Yoshida, M., Kashikawa, N., \& Okamura, S.  2005, \pasj, 57, 447
\bibitem[Shimasaku et al.(2006)]{Shimasaku06} Shimasaku, K. et al.  2006, \pasj, 58, 313
\bibitem[Shioya et al.(2005)]{Shioya05} Shioya, Y. et al.  2005, \pasj, 57, 569
\bibitem[Somerville et al.(2004)]{Some04} Somerville, R.S., Lee, K., Ferguson, H.C., Gardner, J.P., Moustakas, L.A., Giavalisco, M.  2004, \apjl, 600, 171L
\bibitem[Steidel et al.(1998)]{Steidel98} Steidel, C. C., Adelberger, K. L., Dickinson, M., Giavalisco, M., Pettini, M., \& Kellogg, M. 1998, \apj, 492, 428
\bibitem[Stiavelli et al.(2005)]{Stiavelli05} Stiavelli, M. et al.  2005, \apj, 622, L1
\bibitem[Venemans et al.(2002)]{Venemans02} Venemans, B.P. et al.  2002, \apj, 569, L11
\bibitem[Venemans et al.(2005)]{Venemans05} Venemans, B.P. et al.  2005, \aap, 431, 793
\bibitem[Venemans et al.(2007)]{Venemans07} Venemans, B.P. et al.  2007, \aap, 461, 823
\bibitem[Willott et al.(2005)]{Willott05} Willott, C.J., Percival, W.J., McLure, R.J., Crampton, D., Hutchings, J.B., Jarvis, M.J., Sawicki, M., \& Simard, L.  2005, \apj, 626, 657
\bibitem[Yan et al.(2004)]{Yan04} Yan, H. et al.  2004, \apj, 616, 63
\bibitem[Yan \& Windhorst~(2004)]{YanWin04} Yan, H., \& Windhorst, R.A.  2004, \apj, 600, L1
\bibitem[Zheng et al.(2006)]{Zheng06} Zheng, W. et al.  2006, \apj, 640, 574
\end{thebibliography}
\end{document}